\documentclass[prb,twocolumn,preprintnumbers,amsmath,amssymb]{revtex4-1}
\usepackage{amsmath}
\usepackage{amssymb}
\usepackage{mathrsfs}
\usepackage{graphicx}
\usepackage{dcolumn}
\usepackage{bm}
\usepackage{color}
\usepackage{hhline}
\usepackage{rotating}

\begin{document}

\title{High-resolution tunneling spin transport characteristics of topologically distinct magnetic skyrmionic textures from theoretical calculations}

\author{Kriszti\'an Palot\'as$^{1,2,3,4}$}
\email{palotas@phy.bme.hu, palotas.krisztian@wigner.hu}
\author{Levente R\'ozsa$^{5,6}$}
\author{Eszter Simon$^{3,7}$}
\author{L\'aszl\'o Szunyogh$^{3,8}$}

\affiliation{1 Wigner Research Center for Physics, Institute for Solid State Physics and Optics, P.\ O.\ Box 49, H-1525 Budapest, Hungary\\
2 University of Szeged, MTA-SZTE Reaction Kinetics and Surface Chemistry Research Group, Rerrich B.\ t\'er 1, H-6720 Szeged, Hungary\\
3 Budapest University of Technology and Economics, Department of Theoretical Physics, Budafoki \'ut 8, H-1111 Budapest, Hungary\\
4 Slovak Academy of Sciences, Institute of Physics, D\'ubravsk\'a cesta 9, SK-84511 Bratislava, Slovakia\\
5 University of Hamburg, Department of Physics, D-20355 Hamburg, Germany\\
6 University of Konstanz, Department of Physics, D-78457 Konstanz, Germany\\
7 Ludwig Maximilians University, Department of Chemistry, D-81377 Munich, Germany\\
8 Budapest University of Technology and Economics, MTA-BME Condensed Matter Research Group, Budafoki \'ut 8, H-1111 Budapest, Hungary}

\date{\today}

\begin{abstract}

High-resolution tunneling electron spin transport properties (longitudinal spin current (LSC) and spin transfer torque (STT) maps) of topologically distinct real-space magnetic skyrmionic textures are reported by employing a 3D-WKB combined scalar charge and vector spin transport theory in the framework of spin-polarized scanning tunneling microscopy (SP-STM). For our theoretical investigation metastable skyrmionic spin structures with various topological charges ($Q=-3,-2,-1,0,1,2$) in the (Pt$_{0.95}$Ir$_{0.05}$)/Fe/Pd(111) ultrathin magnetic film are considered. Using an out-of-plane magnetized SP-STM tip it is found that the maps of the LSC vectors acting on the spins of the magnetic textures and all STT vector components exhibit the same topology as the skyrmionic objects. In contrast, an in-plane magnetized tip generally does not result in spin transport vector maps that are topologically equivalent to the underlying spin structure, except for the LSC vectors acting on the spins of the skyrmionic textures at a specific relation between the spin polarizations of the sample and the tip. The magnitudes of the spin transport vector quantities exhibit close relations to charge current SP-STM images irrespectively of the skyrmionic topologies. Moreover, we find that the STT efficiency (torque/current ratio) acting on the spins of the skyrmions can reach large values up to $\sim$25 meV/$\mu$A ($\sim$0.97 $h/e$) above the rim of the magnetic objects, but it considerably varies between large and small values depending on the lateral position of the SP-STM tip above the topological spin textures. A simple expression for the STT efficiency is introduced to explain its variation. Our calculated spin transport vectors can be used for the investigation of spin-polarized tunneling-current-induced spin dynamics of topologically distinct surface magnetic skyrmionic textures.


\end{abstract}

\maketitle

\section{Introduction}
\label{sec_int}

Individual magnetic skyrmions are promising building blocks for high-density information storage and low-power information carrier applications in spintronics \cite{nagaosa13,fert13,zhang15,wiesendanger16natrevmat,fert17natrevmat,zhang20,back20} due to their small size and topological properties. Skyrmions and other topological spin textures are formed of closed magnetic domain walls of diverse complexity in real space \cite{Okubo,Leonov,Lin,zhang16scirep,zhang16prb,rozsa-sk3,yang17mobius,hagemeister18kpi_sk,stavrou19,capic19prb,villalba19}. To describe this complexity they are characterized by topological invariants: the winding number or topological charge \cite{nagaosa13,Leonov,leonov16} of the three-dimensional spin vectors, or the vorticity \cite{nagaosa13,Lin,rozsa-sk3,palotas17prb} of their in-plane spin components. First-principles calculations provided deep insight into the role of antisymmetric Dzyaloshinsky-Moriya and isotropic Heisenberg magnetic exchange interactions, exchange frustration, and magnetocrystalline anisotropy in the formation of skyrmions in thin magnetic films \cite{rozsa-sk3,heinze11,dupe14,simon14,rozsa-sk2,malottki17,hsu18,haldar18,meyer19}. Recently, the investigation of the role of higher-order magnetic interactions \cite{romming18,kronlein18,laszloffy19,brinker19} on the skyrmion stability \cite{paul19} is attracting interest. Moving towards realistic spintronic applications several works reported on the existence of stable skyrmions at room temperature \cite{moreau16,boulle16,woo16,soumyanarayanan17,legrand20}. Their stability depending on the temperature (mostly using minimum-energy-path-based methods) and other factors, like the magnetic-field-dependent size and shape, is under active research nowadays \cite{hagemeister15,rozsa-sk1,bessarab15,lobanov16,stosic17,bessarab18,bottcher18,malottki19,bessarab19,desplat19,garanin20,varentcova20}.

For the practical use in future spintronic devices, important aspects of skyrmionic bits are the well-controlled writing, reading, deleting, as well as the movement of the topological spin textures. Tailoring these aspects at will requires an advanced understanding of their spin transport, spin dynamics and spin switching properties. Due to the small size of the skyrmions a viable route to controllably manipulate their spins is through local perturbations either by a focused laser beam, see e.g., Ref.~\onlinecite{polyakov20}, or by the tip field of a magnetic force microscope \cite{vetrova20}, or in an even more localized fashion by using other scanning probe methods with atomically sharp tips.

Spin-polarized scanning tunneling microscopy (SP-STM) has been extensively used to image and manipulate magnetic skyrmions and other complex magnetic objects at surfaces \cite{wiesendanger16natrevmat,palotas17prb,romming15prl,bergmann14,wiesendanger09review}. Controlled creation and annihilation of skyrmions have been reported in the seminal work of Romming et al.~\cite{romming13} using local current pulses of the tip of an STM with opposite voltage polarities. These effects have been explained by the electric fields in the STM junction \cite{hsu17elec}, and very recently the transition rate was mapped on the nanometer scale \cite{muckel20}, but the roles of the tunneling spin transfer torque (STT) and other spin transport processes in an SP-STM are less understood, and only a few works addressed such questions so far \cite{wieser17,palotas16prb,palotas18stt1,palotas18stt2}. Wieser et al.~\cite{wieser17} demonstrated the creation, movement, and annihilation of a skyrmion with an SP-STM tip by theoretical means. Their employed STT vectors in the Landau-Lifshitz-Gilbert (LLG) equation of spin dynamics are calculated by assuming maximally spin-polarized tips ($P_T=\pm1$) within the Tersoff-Hamann approximation \cite{tersoff83,tersoff85} of electron tunneling. An electron tunneling model capable of describing the scalar charge current and vector spin transport in a consistent way has been proposed in our previous works \cite{palotas16prb,palotas18stt1,palotas18stt2}, where the electronic structures of the sample and the tip based on first-principles calculations can be incorporated \cite{palotas16prb}, going beyond the Tersoff-Hamann approximation.

In the present work, the tunneling electron spin transport properties of six topologically distinct magnetic skyrmions in an ultrathin film are reported based on theoretical calculations. We utilize an electron tunneling theory for the combined calculation of scalar charge and vector spin transport in SP-STM within the three-dimensional (3D) Wentzel-Kramers-Brillouin (WKB) framework \cite{palotas16prb,palotas18stt1,palotas18stt2}. The topologies of the maps of the calculated tunneling vector spin transport quantities, the longitudinal spin current (LSC) and the STT, are compared with those of the underlying spin structures, depending on the magnetization orientation of the SP-STM tip. The magnitudes of the spin transport LSC and STT vector quantities exhibit close relations to charge current SP-STM images irrespectively of the skyrmionic topologies. An important quantity, the STT efficiency, measures the exerted torque on the spins of the skyrmionic structures per unit charge current \cite{palotas16prb}. Such STT efficiency maps are reported in high spatial resolution for the first time above topologically different skyrmionic spin textures. We find a great variation of the STT efficiency depending on the lateral position of the SP-STM tip, and we identify regions for large values up to $\sim$25 meV/$\mu$A ($\sim$0.97 $h/e$) above the rim of the magnetic objects.

The paper is organized as follows. Together with some general considerations, the studied skyrmionic spin structures with different topologies are briefly described in section \ref{sec_sky}. In section \ref{sec_meth} the combined tunneling electron charge and vector spin transport 3D-WKB theoretical model in SP-STM considering noncollinear magnetic surfaces is briefly presented. The tunneling vector spin transport (LSC and STT) properties of the topologically distinct skyrmionic textures and the relations to the electronic charge current, in particular the STT efficiency, are reported in section \ref{sec_res}. Summary and conclusions are found in section \ref{sec_conc}.

\section{Skyrmionic spin structures}
\label{sec_sky}

\begin{figure*}[t]
\includegraphics[width=1.0\textwidth,angle=0]{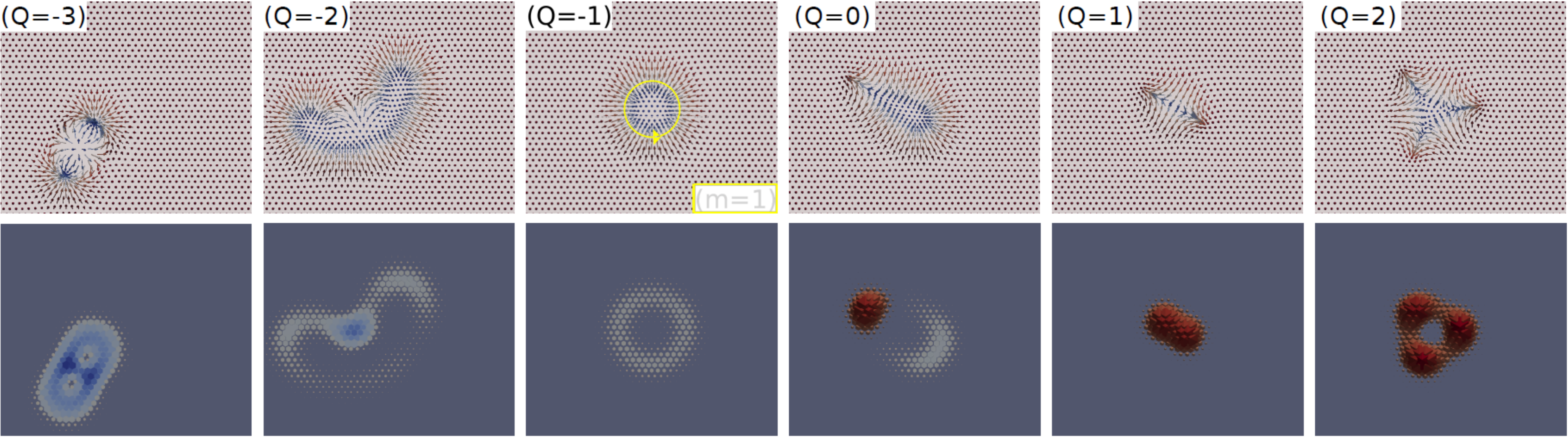}
\caption{\label{Fig1} Skyrmionic spin structures (data taken from Ref.~\onlinecite{rozsa-sk3}) with various topological charges $Q$ (first row), and the emergent magnetic field vectors, $\mathbf{B}_\mathrm{eff}^a$, at atomic positions "$a$" due to the topological charge density $q_a$ according to Eq.~(\ref{Eq_field}) (second row). Red and blue colors respectively correspond to positive and negative out-of-plane ($z$) vector components: spin (first row) and $\mathbf{B}_\mathrm{eff}^a$ (second row). For determining the vorticity ($m$) an illustrative example is shown for $m=1$ $(Q=-1)$ by a yellow circular arrow as a closed curve, along which the number of counterclockwise $360^{\circ}$ rotations of the in-plane (gray) spins has to be counted.}
\end{figure*}

In a continuum description the classical spin configurations of topological magnetic objects (e.g. skyrmions) can be represented by a vector field of unit length, $\mathbf{s}_S(\mathbf{r})$, which has a spatial dependence in the two-dimensional (2D) surface (denoted by the subscript $S$) plane. The winding number or topological charge $Q$ of the vector field counts the number of times $\mathbf{s}_S(\mathbf{r})$ winds around the unit sphere:
\begin{equation}
Q=\int q(\mathbf{r})d^2r=\frac{e}{h}\int B_{\mathrm{eff}}(\mathbf{r})d^2r,
\label{Eq_Q}
\end{equation}
where the integrals are performed over the surface, $h$ is Planck's constant, $e$ is the elementary charge, $h/e$ is the magnetic flux quantum, and $B_{\mathrm{eff}}(\mathbf{r})=(h/e)q(\mathbf{r})$ is the $z$-component of the emergent magnetic field \cite{palotas18stt2,raju19} due to the topological charge density $q(\mathbf{r})=\mathbf{s}_S(\mathbf{r})\cdot(\partial_x\mathbf{s}_S(\mathbf{r})\times\partial_y\mathbf{s}_S(\mathbf{r}))/4\pi$ of the real-space spin texture.

For our analysis it is important to introduce another topological invariant, the vorticity \cite{nagaosa13,Lin} of the spin texture. The vorticity ($m$) counts the number of times the in-plane components of $\mathbf{s}_S(\mathbf{r})$ rotate around the circle when following an arbitrary closed curve in the surface plane enclosing the center of the localized spin texture \cite{rozsa-sk3,palotas17prb}. The sign of $m$ is characteristic for the direction of such an in-plane spin rotation, and the best choice for the closed curve is above the largest in-plane components of the spin structures for a visual analysis. Note that for the skyrmionic spin structures investigated here, $Q$ is related to $m$ by the spin direction of the out-of-plane ferromagnetic background far from the localized skyrmionic texture $\mathbf{s}_S(|\mathbf{r}|\rightarrow\infty)$ as \cite{rozsa-sk3}: $Q=-m\mathbf{e}_z\cdot\mathbf{s}_S(|\mathbf{r}|\rightarrow\infty)$.

The continuum description of the topological magnetic configurations consisting of classical spins can be transformed to discrete lattices, where the spin vector field of unit length has an atomic site "$a$"-dependence, $\mathbf{s}_S^a$, in the 2D surface plane. Recipes for calculating the scalar chirality of three-spin-plaquettes of $\mathbf{s}_S^a$ on 2D lattices can be found, e.g., in Refs.~\cite{rozsa-sk1,berg81,yin16}. The scalar chirality can also be defined as an atomic-site-dependent quantity \cite{palotas18stt2,fernandes20}:
\begin{equation}
C_a=\frac{1}{n\pi}\sum_{i,j}\arctan\left[{\frac{\mathbf{s}_S^a\cdot(\mathbf{s}_S^i\times\mathbf{s}_S^j)}{1+\mathbf{s}_S^a\cdot\mathbf{s}_S^i+\mathbf{s}_S^a\cdot\mathbf{s}_S^j+\mathbf{s}_S^i\cdot\mathbf{s}_S^j}}\right],
\label{Eq_chirality}
\end{equation}
where "$i$" and "$j$" are among the neighboring spins of site "$a$", and \emph{all} (number of $n$) triangular plaquettes are formed with an "$a-i-j$" counterclockwise order with the usual choice of Cartesian axes and looking toward the $-z$ direction as in all figures reported in this paper. Based on $C_a$ the discretized topological charge density is
\begin{equation}
q_a=C_a/A_0,
\end{equation}
where $A_0$ is the surface area associated with each atomic site in the 2D lattice of the surface plane. Due to the topological charge density $q_a$, there is an emergent magnetic field at site "$a$": $\mathbf{B}_\mathrm{eff}^a=B_\mathrm{eff}^a\mathbf{e}_z$, where $B_\mathrm{eff}^a$ is the discrete equivalent of $B_{\mathrm{eff}}(\mathbf{r})$ in Eq.~(\ref{Eq_Q}), and
\begin{equation}
\mathbf{B}_\mathrm{eff}^a=(h/e)q_a\mathbf{e}_z=B_0 C_a\mathbf{e}_z,
\label{Eq_field}
\end{equation}
where $B_0=h/(eA_0)$ is the unit of the emergent magnetic field (area density of the magnetic flux quantum) characteristic for the 2D lattice. With all of these, the topological charge $Q$ in Eq.~(\ref{Eq_Q}) can be calculated on the discrete 2D lattice as
\begin{equation}
Q=\sum_a C_a=A_0\sum_a q_a=\sum_a B_\mathrm{eff}^a/B_0.
\label{Eq_Q_discrete}
\end{equation}

The considered six topologically distinct metastable skyrmionic objects - spin structures on a discrete 2D hexagonal lattice of atomic sites - under investigation in the present paper were obtained by a combination of {\it{ab initio}} and spin dynamics calculations in the (Pt$_{0.95}$Ir$_{0.05}$)Fe/Pd(111) ultrathin magnetic film, and their stability was analyzed in detail \cite{rozsa-sk3,rozsa-sk2}. The resulting spin structures, which are expected to remain stable at temperature and external field values achievable in low-temperature SP-STM experiments \cite{rozsa-sk3}, are shown in the first row of Figure \ref{Fig1}. For their topological characterization, the scalar chirality of the spin at site "$a$" on a hexagonal lattice is obtained by using Eq.~(\ref{Eq_chirality}) with $n=6$. Based on the calculated $C_a$ values, the discrete topological charge density $q_a$, and following Eq.~(\ref{Eq_Q_discrete}) the topological charge ($Q$) of the spin structures in Fig.~\ref{Fig1} can be determined: they range from $Q=-3$ to $Q=2$ as shown in Fig.~\ref{Fig1}. The vorticity ($m=-Q$ in Fig.~\ref{Fig1}) can be obtained by counting the number of counterclockwise $360^{\circ}$ rotations of the in-plane spin components along the perimeter of the spin structures. An illustrative example is shown by a yellow circular arrow as a closed curve above the largest in-plane (gray) spin components for $m=1$ $(Q=-1)$ in Fig.~\ref{Fig1}. The second row of Fig.~\ref{Fig1} reports the site-dependent emergent magnetic field vectors, $\mathbf{B}_\mathrm{eff}^a$, due to the topological charge density $q_a$ following Eq.~(\ref{Eq_field}). Note that the value of $B_0$ in the considered (Pt$_{0.95}$Ir$_{0.05}$)Fe/Pd(111) surface system is $6.3\times10^4$ T.

\section{3D-WKB electron tunneling theory}
\label{sec_meth}

The tunneling electron charge and spin transport properties of the magnetic skyrmions are described within the 3D-WKB electron tunneling theory \cite{palotas17prb,palotas16prb,palotas18stt1,palotas18stt2,palotas11sts,palotas11stm,palotas12sts,palotas12orb,palotas13contrast,mandi13tiprot,nita14,mandi14fe,mandi14rothopg,mandi15tipstat}, and the theoretical calculations are performed using the 3D-WKB-STM code \cite{palotas14fop}. The tunneling transport properties are determined at the tip apex position, $\mathbf{R}_{T}$, by calculating a superposition (sum over "$a$") of 1D WKB tunneling electron transition contributions between the magnetic tip apex atom (characterized by a spin unit vector $\mathbf{s}_T$) and the surface atoms "$a$" at positions $\mathbf{R}_a$ (characterized by local spin unit vectors $\mathbf{s}_S^a$). The scalar charge current ($I$) \cite{heinze06}, the out-of-plane $\mathbf{T}^{\perp}$ and in-plane $\mathbf{T}^{j\parallel}$ ($j\in\{T,S\}$) components of the STT vectors, and the LSC vectors $\mathbf{T}^{jL}$ ($j\in\{T,S\}$) at $\mathbf{R}_{T}$ are calculated in the limits of elastic tunneling and low bias voltage $V$ as \cite{palotas18stt1}
\begin{eqnarray}
I(\mathbf{R}_{T},V)&=&\frac{e^2}{h}|V|\sum_a t(\mathbf{R}_{T}-\mathbf{R}_a)(1+P_{S}P_{T}\cos\phi_a),\nonumber\\
\mathbf{T}^{\perp}(\mathbf{R}_{T},V)&=&e|V|\sum_a t(\mathbf{R}_{T}-\mathbf{R}_a)P_{S}P_{T}\mathbf{s}_S^a\times\mathbf{s}_T,\nonumber\\
\mathbf{T}^{T\parallel}(\mathbf{R}_{T},V)&=&eV\sum_a t(\mathbf{R}_{T}-\mathbf{R}_a)P_{S}\mathbf{s}_T\times(\mathbf{s}_S^a\times\mathbf{s}_T),\nonumber\\
\mathbf{T}^{S\parallel}(\mathbf{R}_{T},V)&=&eV\sum_a t(\mathbf{R}_{T}-\mathbf{R}_a)P_{T}\mathbf{s}_S^a\times(\mathbf{s}_T\times\mathbf{s}_S^a),\nonumber\\
\mathbf{T}^{TL}(\mathbf{R}_{T},V)&=&eV\sum_a t(\mathbf{R}_{T}-\mathbf{R}_a)(P_{T}+P_{S}\cos\phi_a)\mathbf{s}_T,\nonumber\\
\mathbf{T}^{SL}(\mathbf{R}_{T},V)&=&eV\sum_a t(\mathbf{R}_{T}-\mathbf{R}_a)(P_{S}+P_{T}\cos\phi_a)\mathbf{s}_S^a.\nonumber\\
\label{Eq_transport}
\end{eqnarray}
Here, $\phi_a$ is the angle between the spin moment of surface atom "$a$" and the tip apex atom, thus $\cos\phi_a=\mathbf{s}_S^a\cdot\mathbf{s}_T$. $P_S$ and $P_T$ denote the scalar spin polarization of the surface atoms and the tip apex atom at their respective Fermi energies, and they are independent parameters in our tunneling model \cite{palotas18stt1}. In the present work $P_S=-0.5$ and $P_T=-0.8$ spin polarization values were selected, and $|V|=1.5$ meV absolute bias voltage has been considered. Equation (\ref{Eq_transport}) enables the calculation of the tunneling electron charge and spin transport quantities with an arbitrarily high spatial resolution by adjusting the tip position $\mathbf{R}_{T}$, approaching the continuum limit where topological arguments become applicable.

The electron transmission function is \cite{heinze06}
\begin{equation}
t(\mathbf{r})=\exp\left[-4\pi\sqrt{2M\Phi}|\mathbf{r}|/h\right],
\label{Eq_transmission}
\end{equation}
where $M$ is the mass of the electron, and $\Phi$ is the effective work function. In the present work $\Phi=5$ eV is selected. In the transmission function $t(\mathbf{r})$ all electronic states are assumed to be exponentially decaying spherical states \cite{tersoff83,tersoff85,heinze06}, and the electron-orbital dependence \cite{palotas16prb,palotas12orb,mandi13tiprot,mandi14fe,mandi14rothopg,mandi15tipstat} of $t(\mathbf{r})$ is omitted for simplicity. The latter would play a significant role at higher bias voltages with realistic composition of the densities of states for all electron orbitals involved in the tunneling. Such a functionality is implemented in the 3D-WKB-STM code, and could be employed in the future in combination with first-principles methods when direct comparison with high-resolution spin transport SP-STM measurements would be seeked. Due to the fast decay of $t(\mathbf{r})$, in the following discussion the spin direction $\mathbf{s}_S^A$ of the surface atom $A$, which is closest to the tip apex position $\mathbf{R}_{T}$ (below the tip), is understood when referring to a single $\phi_A$ value \cite{palotas18stt1}.

The upper indices $j\in\{T,S\}$ in $\mathbf{T}^{j\parallel}$ and $\mathbf{T}^{jL}$ in Eq.~(\ref{Eq_transport}) denote the tip or sample side, on which spin moments the tunneling in-plane STT and the LSC are acting. With this distinction, the total STT vectors can be obtained as \cite{palotas18stt1}
\begin{eqnarray}
\mathbf{T}^{T}(\mathbf{R}_{T},V)&=&\mathbf{T}^{\perp}(\mathbf{R}_{T},V)+\mathbf{T}^{T\parallel}(\mathbf{R}_{T},V)\\
\mathbf{T}^{S}(\mathbf{R}_{T},V)&=&\mathbf{T}^{\perp}(\mathbf{R}_{T},V)+\mathbf{T}^{S\parallel}(\mathbf{R}_{T},V).
\end{eqnarray}
The STT efficiency maps acting on the surface atomic spins are calculated as
\begin{equation}
\eta(\mathbf{R}_{T},V)=|\mathbf{T}^{S}(\mathbf{R}_{T},V)|/I(\mathbf{R}_{T},V).
\label{Eq_STT_eff}
\end{equation}
A simple expression for the STT efficiency is proposed in Eq.~(\ref{Eq_STT_eff_A}) in section \ref{sec_res} to understand its governing factors.

Note that the visualizational parameters for the presentation of the results in the present work correspond to those employed in Refs.~\cite{palotas17prb,palotas18stt1,palotas18stt2}.

\section{Results and discussion}
\label{sec_res}

Employing the above-described combined electron tunneling charge and vector spin transport 3D-WKB theory, a set of skyrmionic spin structures with different topological charges (shown in Fig.~\ref{Fig1}) is considered, and their spin transport properties are investigated in high spatial resolution, also in relation to charge transport properties.

\begin{figure*}[t]
\includegraphics[width=1.0\textwidth,angle=0]{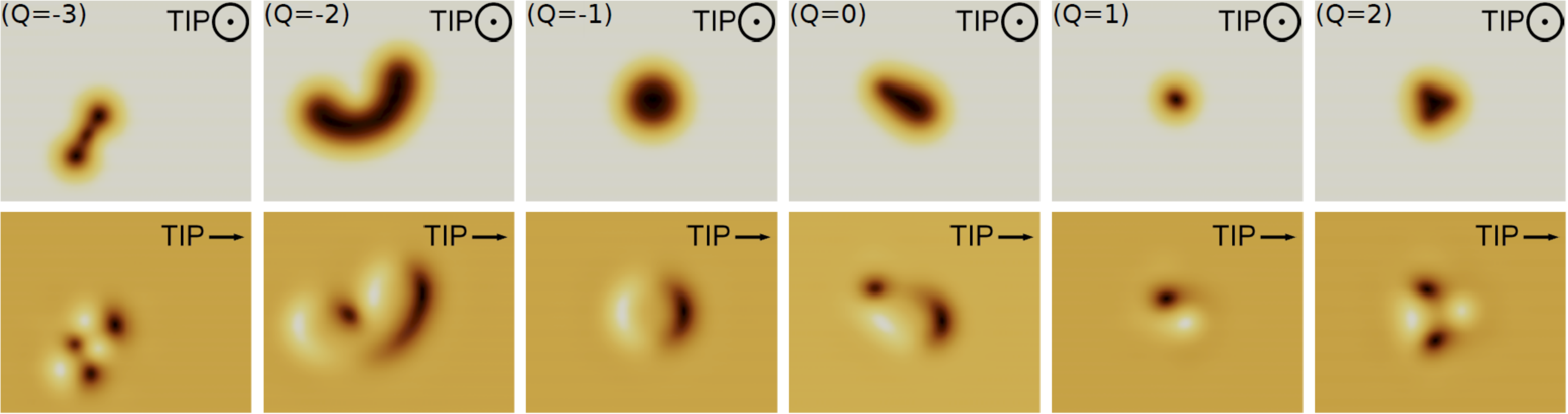}
\caption{\label{Fig2} Constant-current SP-STM images (taken from Ref.~\onlinecite{palotas17prb}) of the skyrmionic spin structures in Fig.~\ref{Fig1} at $|V|=1.5$ meV using two differently oriented magnetic tips: an out-of-plane (pointing to the $+z$ $[111]$ direction) and an in-plane (pointing to the $+x$ $[1\bar{1}0]$ direction). The tip magnetization orientations are explicitly indicated. Bright and dark contrast respectively means higher and lower apparent height of the constant-current contour.}
\end{figure*}

Figure \ref{Fig2} shows a set of charge current SP-STM images \cite{palotas17prb} of the different skyrmionic objects listed in Fig.~\ref{Fig1}. The images in the first row of Fig.~\ref{Fig2} are taken with an out-of-plane magnetized tip, and diverse contrasts can be identified, which resemble the shape of the skyrmions at first sight (compare with Fig.~\ref{Fig1}), and more rigorously they correspond to the symmetries of the spin structures \cite{palotas17prb}: axial symmetry for $Q=-1$, and $C_{|1+Q|}$ symmetry for $Q\ne-1$. The second row of Fig.~\ref{Fig2} shows SP-STM images with an in-plane magnetized tip. Except for the skyrmionic object with $Q=0$ (chimera skyrmion), the number of bright and dark contrast regions each corresponds to the absolute value of Q for all spin structures \cite{palotas17prb}: a skyrmion and an antiskyrmion ($|Q|=1$), a double skyrmion and a double antiskyrmion ($|Q|=2$), and a triple skyrmion ($|Q|=3$). A more detailed discussion on the SP-STM contrasts, their rotation with respect to in-plane tip magnetization rotations, and relation to the topological charge density can be found in Ref.~\onlinecite{palotas17prb}.

\begin{figure*}[t]
\includegraphics[width=1.0\textwidth,angle=0]{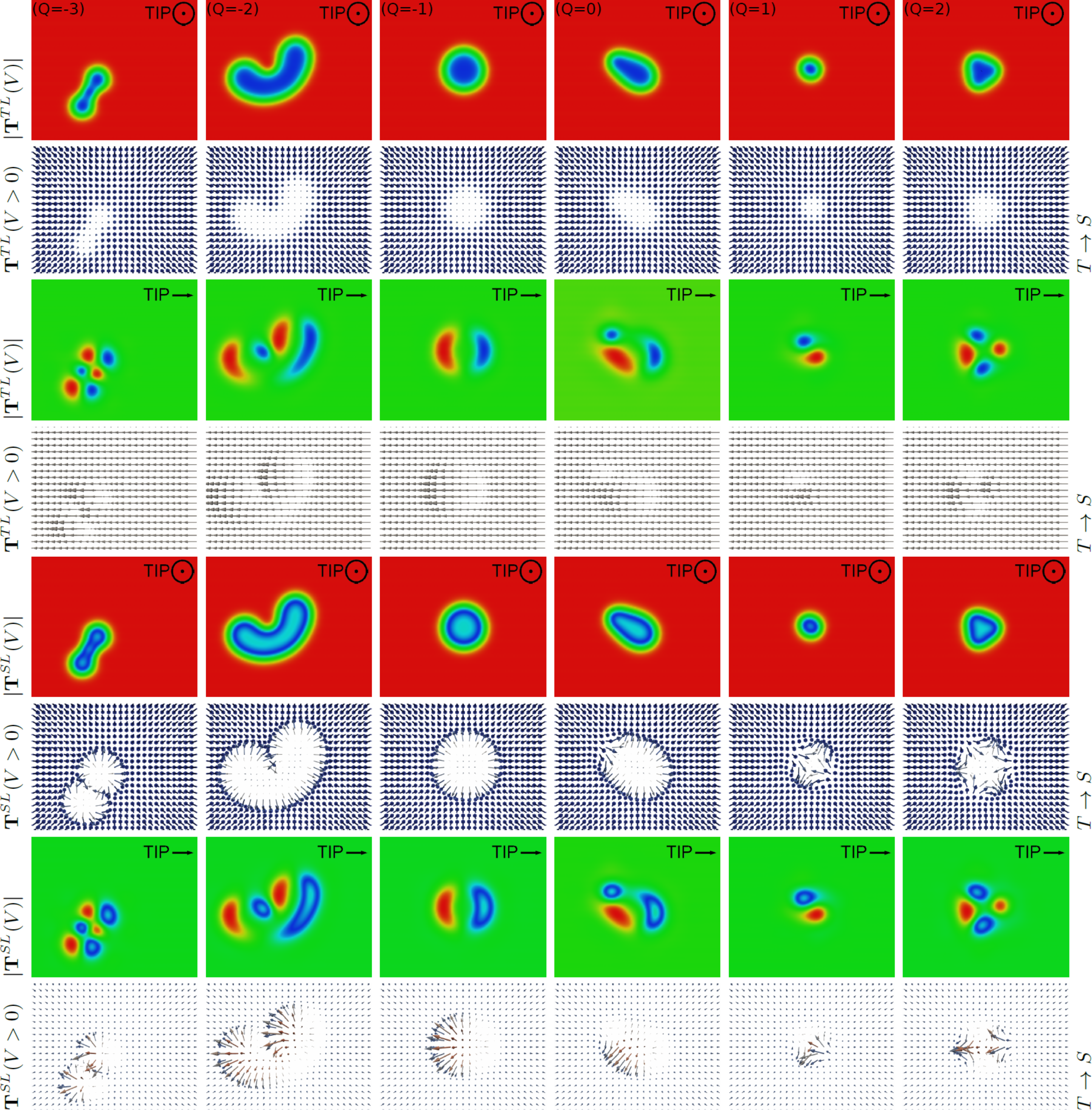}
\caption{\label{Fig3} Longitudinal spin current (LSC) magnitudes and vectors acting on the scanning tip, $|\mathbf{T}^{TL}(V)|$ and $\mathbf{T}^{TL}(V)$, and on the skyrmionic spin structures, $|\mathbf{T}^{SL}(V)|$ and $\mathbf{T}^{SL}(V)$, 6 \AA\;above the magnetic textures shown in Fig.~\ref{Fig1} at $|V|=1.5$ meV using an out-of-plane and an in-plane magnetized tip (magnetization directions are explicitly indicated). The LSC vectors are given at positive bias voltage ($V>0$), i.e., at $T\rightarrow S$ tunneling direction, and their red and blue colors correspond to positive and negative out-of-plane ($z$) vector components, respectively. The color scales of the LSC magnitudes correspond to $|\mathbf{T}^{TL}(+z)|$ (first row): red maximum at 5.7 neV, blue minimum at 1.3 neV; $|\mathbf{T}^{TL}(+x)|$ (third row): red maximum at 5.6 neV, blue minimum at 1.4 neV; $|\mathbf{T}^{SL}(+z)|$ (fifth row): red maximum at 5.7 neV, blue minimum at 0.2 neV; and $|\mathbf{T}^{SL}(+x)|$ (seventh row): red maximum at 5.2 neV, blue minimum at 0 neV.}
\end{figure*}

In the following, we focus on the tunneling spin transport properties of the skyrmionic objects with various topologies shown in Fig.~\ref{Fig1}. Figure \ref{Fig3} reports their 2D maps of calculated longitudinal spin current (LSC) magnitudes and vectors obtained at a constant-height condition by a scanning magnetic tip. An out-of-plane and an in-plane magnetized tip is considered, and their magnetization directions are shown in the images of the LSC magnitudes. The top and bottom half of Fig.~\ref{Fig3} contain LSC data acting on the spin of the tip apex atom ($\mathbf{T}^{TL}$) and on the spins of the skyrmionic objects on the sample surface ($\mathbf{T}^{SL}$), respectively. When comparing with Fig.~\ref{Fig2} we find that the contrast patterns of the LSC magnitudes qualitatively correspond to those of the charge current for all skyrmionic topologies. The reasons for their apparently similar $\phi_A$-dependence can be deduced from Eq.~(\ref{Eq_transport}) and they were analyzed in detail for the $Q=-1$ skyrmion in Ref.~\onlinecite{palotas18stt1}. Fig.~\ref{Fig3} clearly demonstrates that the identified relation between the LSC magnitude and the charge current is not affected by the topology of the skyrmionic spin structure. Thus, we propose that the tip-position-dependent contrast of the LSC magnitudes can be qualitatively predicted based on the known (measured or calculated) charge current SP-STM images of skyrmionic objects with an arbitrary topology of their real-space spins.

The 2D vector maps of Fig.~\ref{Fig3} show the LSC vectors $\mathbf{T}^{TL}$ and $\mathbf{T}^{SL}$ at positive bias voltage ($V>0$), i.e., at $T\rightarrow S$ tunneling direction. Due to the definitions of the LSC vectors in Eq.~(\ref{Eq_transport}), changing the sign of the bias voltage to $V<0$, thus the tunneling direction to $S\rightarrow T$, would result in LSC vectors of opposite directions each. The $\mathbf{T}^{TL}$ vectors are always in line with the spin direction of the tip apex atom ($\mathbf{s}_T$), and their parallel or antiparallel alignment is determined by the sign of $P_T$ and the relation of $|P_T|$ to $|P_S|$, generally by the sign of $(P_{T}+P_{S}\cos\phi_A)$ \cite{palotas18stt1}. As can be seen in the top half of Fig.~\ref{Fig3}, the $\mathbf{T}^{TL}$ vectors point oppositely to $\mathbf{s}_T$ due to $P_T=-0.8$, and the vector magnitudes clearly reflect the scalar $|\mathbf{T}^{TL}|$ magnitudes depending on the tip position. Similarly, the $\mathbf{T}^{SL}$ vectors are always in line with the spin direction of the surface atom closest to the tip position ($\mathbf{s}_S^A$), and their parallel or antiparallel orientation is determined by the sign of $P_S$ and the relation of $|P_S|$ to $|P_T|$, generally by the sign of $(P_{S}+P_{T}\cos\phi_A)$ \cite{palotas18stt1}. Comparing with the spin structures in Fig.~\ref{Fig1}, as can be seen in the bottom half of Fig.~\ref{Fig3}, the $\mathbf{T}^{SL}$ vectors generally point oppositely to $\mathbf{s}_S^A$ due to $P_S=-0.5$, except for the regions with small $|\mathbf{T}^{SL}|$ magnitudes if $|P_T|>|P_S|$ (as in the presented case), that is in the regions enclosed by the $|\mathbf{T}^{SL}|$ minima shown as blue belts in Fig.~\ref{Fig3}, where $\cos\phi_A<-P_S/P_T$ ($-0.625$ in the presented case) \cite{palotas18stt1}.

Concerning the topological properties of the calculated (normalized) LSC vector maps, the following statements can be made. The $\mathbf{T}^{TL}$ vectors do not show topological properties due to their in-line direction with the spin of the tip apex atom. Since the $\mathbf{T}^{SL}$ vectors generally follow the direction of $\mathbf{s}_S^A$ by the scanning tip if $P_S>0$ and the opposite direction of $\mathbf{s}_S^A$ if $P_S<0$, they are good candidates to exhibit the same topological properties as the underlying spin structure $\mathbf{s}_S^a$. These, however, depend on the relation between the spin polarizations of the sample and the tip. The case of $|P_T|>|P_S|$ is shown in Fig.~\ref{Fig3}, where the out-of-plane magnetized tip results in $\mathbf{T}^{SL}$ vectors that clearly show the same topology as $\mathbf{s}_S^a$ (Fig.~\ref{Fig1}) in the sense of the vorticity but not the topological charge. An in-plane magnetized tip does not provide topological correspondence between the $\mathbf{T}^{SL}$ and $\mathbf{s}_S^a$ vectors neither in the sense of the vorticity nor of the topological charge, due to the presence of the small regions exhibiting reversed $\mathbf{T}^{SL}$ vectors compared to the general trend (within the blue belts of $|\mathbf{T}^{SL}|$ minima in Fig.~\ref{Fig3}). Such regions with reversed $\mathbf{T}^{SL}$ vectors are completely non-existing if $|P_T|<|P_S|$, and in this case both out-of-plane and in-plane magnetized tips result in $\mathbf{T}^{SL}$ vectors that are topologically equivalent to $\mathbf{s}_S^a$ in the sense of both the vorticity and the topological charge (not shown here). Such examples for the $Q=-1$ skyrmion can be seen in Fig.~5 of Ref.~\onlinecite{palotas18stt1}.

\begin{figure*}[t]
\includegraphics[width=1.0\textwidth,angle=0]{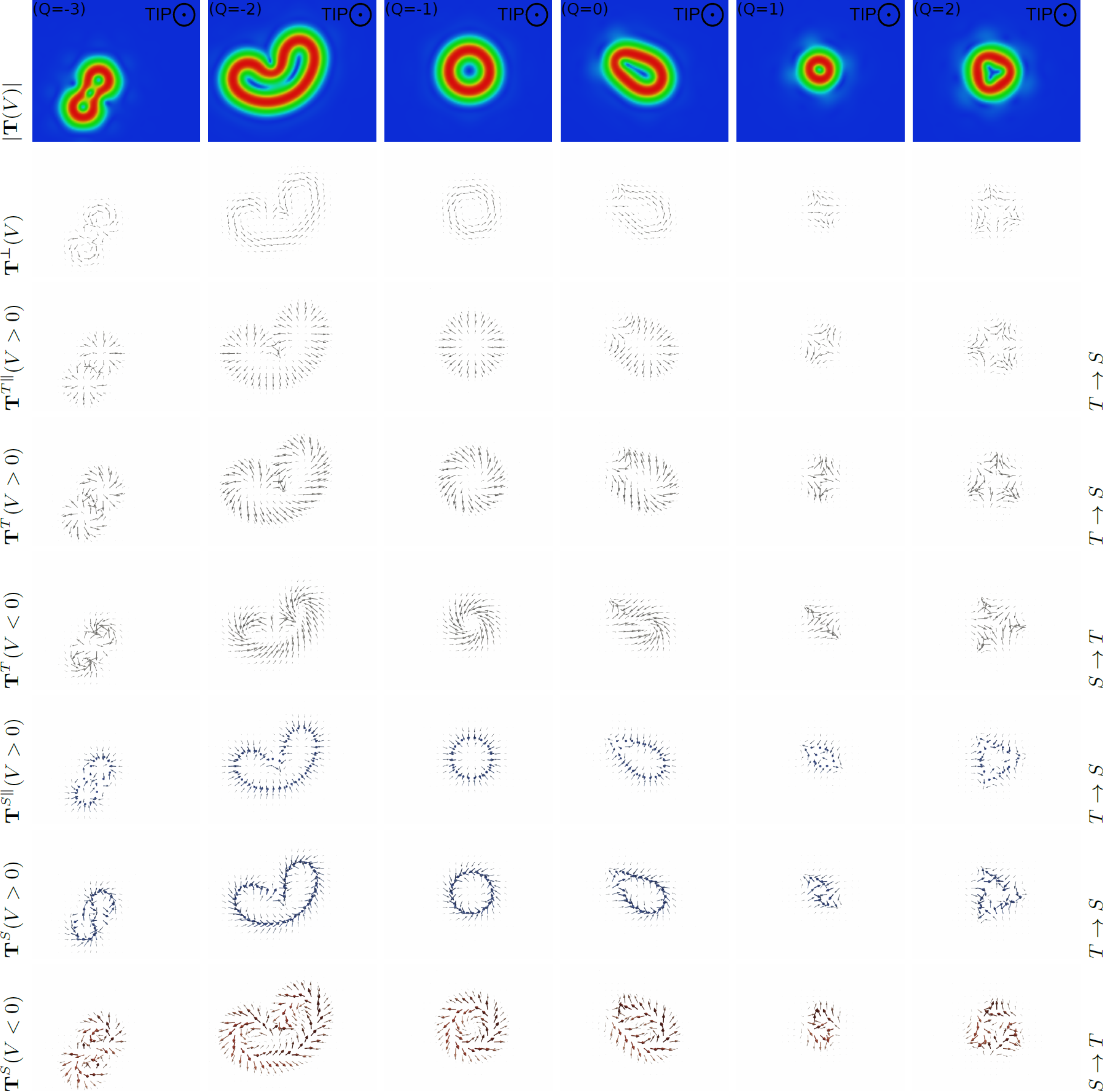}
\caption{\label{Fig4} Spin transfer torque (STT) magnitudes $|\mathbf{T}(V)|$ and vectors [out-of-plane component $\mathbf{T}^{\perp}(V)$, in-plane component $\mathbf{T}^{j\parallel}(V)$, total $\mathbf{T}^{j}(V)=\mathbf{T}^{\perp}(V)+\mathbf{T}^{j\parallel}(V)$] acting on the spin of the scanning tip apex atom ($j=T$) and on the spins of the skyrmionic textures ($j=S$) shown in Fig.~\ref{Fig1} at $|V|=1.5$ meV using an out-of-plane ($+z=[111]$ direction) magnetized tip at 6 \AA\;tip-sample distance. The $\mathbf{T}^{\perp}(V)$ vectors are the same for both tunneling directions, the $\mathbf{T}^{j\parallel}(V)$ vectors are shown for positive bias voltage ($V>0$), i.e., at $T\rightarrow S$ tunneling direction, and the total $\mathbf{T}^{j}(V)$ vectors are given at both bias polarities and tunneling directions. Red and blue colors of the STT vectors correspond to positive and negative out-of-plane ($z$) vector components, respectively. The color scales of the STT magnitudes correspond to $|\mathbf{T}^{T}(+z)|$: red maximum at 2.7 neV, blue minimum at 0 neV; and $|\mathbf{T}^{S}(+z)|$: red maximum at 3.6 neV, blue minimum at 0 neV.}
\end{figure*}

Figure \ref{Fig4} shows 2D maps of calculated spin transfer torque (STT) magnitudes, out-of-plane and in-plane STT vector components, and total STT vectors for the topologically distinct skyrmionic spin textures in Fig.~\ref{Fig1} at a constant-height condition, employing an out-of-plane magnetized scanning tip. Since all STT components have a $\sin\phi_A$-dependence \cite{palotas18stt1} due to the vector product $\mathbf{s}_S^A\times\mathbf{s}_T$ in Eq.~(\ref{Eq_transport}), the magnitudes of the STT components are qualitatively similar and their 2D maps are practically the same, and this is denoted by $|\mathbf{T}|$ and shown in the top row of Fig.~\ref{Fig4}. Taking an out-of-plane magnetized tip, the STT minima (blue regions in Fig.~\ref{Fig4}) and maxima (red regions in Fig.~\ref{Fig4}) are observed above surface regions with dominating out-of-plane and in-plane spin components, respectively. The maximal STT (red) regions corresponding to the in-plane spins of the skyrmionic objects follow their shapes and symmetries: axial symmetry for $Q=-1$, and $C_{|1+Q|}$ symmetry for $Q\ne-1$. In relation to the charge current ($\cos\phi_A$-dependence), it was established that the STT minima are observed at the regions, where the charge current has maxima or minima \cite{palotas18stt1}. The top row of Fig.~\ref{Fig4} in relation to the top row of Fig.~\ref{Fig2} clearly demonstrates that the identified relation between the STT magnitude and the charge current is not affected by the topology of the skyrmionic spin structure. Thus, we propose that the tip-position-dependent contrast of the STT magnitudes can be qualitatively predicted based on the known (measured or calculated) charge current SP-STM images of skyrmionic objects with an arbitrary topology of their real-space spins.

Below the row of the STT magnitudes in Fig.~\ref{Fig4}, the 2D maps of the STT vectors and vector components are shown. Note that the $\mathbf{T}^{\perp}$ and the $\mathbf{T}^{T\parallel}$ components, thus, the total STT vectors acting on the spin of the tip apex atom, $\mathbf{T}^{T}$, lie in the surface plane since the tip is magnetized in the out-of-plane direction. On the other hand, the $\mathbf{T}^{S\parallel}$ and $\mathbf{T}^{S}$ vectors do not lie in the surface plane. This can be understood from the fact that the $\mathbf{T}^{S\parallel}$ vectors always lie in the local $\mathbf{s}_S^A-\mathbf{s}_T$ planes (and $\mathbf{T}^{S\parallel}\perp\mathbf{s}_S^A$), which vary in the skyrmionic spin structures depending on the tip position. A more detailed explanation on the STT vectors and their components for the $Q=-1$ skyrmion is given in Ref.~\onlinecite{palotas18stt1}.

It is important to find in Fig.~\ref{Fig4} that the (non-zero) $\mathbf{T}^{\perp}$ vectors exhibit the same topology as the underlying spin structure $\mathbf{s}_S^a$ (see Fig.~\ref{Fig1}) in the sense of the vorticity, i.e., the rotation direction of the in-plane-lying $\mathbf{T}^{\perp}$ vectors along the perimeter of the spin structures corresponds to the vorticity of the skyrmionic texture irrespective of their topological charge value. This finding holds for the (non-zero) in-plane-lying $\mathbf{T}^{T\parallel}(V)$ and $\mathbf{T}^{T}(V)=\mathbf{T}^{\perp}(V)+\mathbf{T}^{T\parallel}(V)$ vectors, and for the (non-zero) in-plane components of the $\mathbf{T}^{S\parallel}(V)$ and $\mathbf{T}^{S}(V)=\mathbf{T}^{\perp}(V)+\mathbf{T}^{S\parallel}(V)$ vectors as well, where the rotational direction of the in-plane components of the listed torque vectors corresponds to the vorticity of the spin texture.

\begin{figure*}[t]
\includegraphics[width=1.0\textwidth,angle=0]{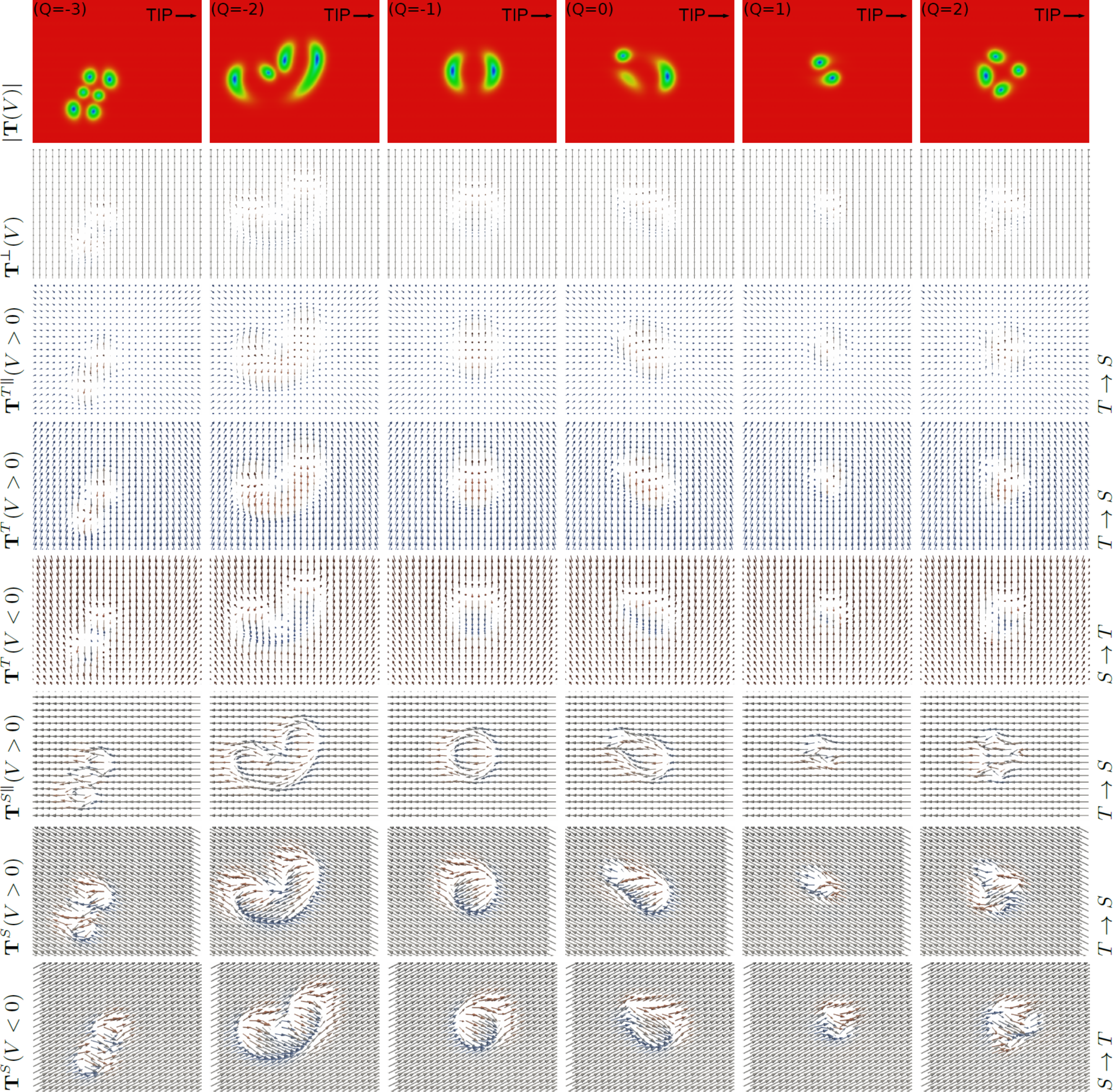}
\caption{\label{Fig5} Same as in Fig.~\ref{Fig4} using an in-plane ($+x=[1\bar{1}0]$ direction) magnetized tip. The color scales of the STT magnitudes correspond to $|\mathbf{T}^{T}(+x)|$: red maximum at 2.8 neV, blue minimum at 0 neV; and $|\mathbf{T}^{S}(+x)|$: red maximum at 3.9 neV, blue minimum at 0.4 neV.}
\end{figure*}

Figure \ref{Fig5} shows 2D maps of calculated STT magnitudes, out-of-plane and in-plane STT vector components, and total STT vectors for the topologically distinct skyrmions in Fig.~\ref{Fig1} at a constant-height condition, employing an in-plane magnetized scanning tip. Again, the 2D maps of the magnitudes of the out-of-plane, in-plane, and total STT vectors for a given skyrmion are essentially the same due to the $\sin\phi_A$-dependence with different scaling factors \cite{palotas18stt1}, and the STT magnitude is denoted by $|\mathbf{T}|$ and shown in the top row of Fig.~\ref{Fig5}. The STT minima and maxima (blue and red regions in Fig.~\ref{Fig5}, respectively) are obtained where the spins of the skyrmions are in line (parallel or antiparallel) with and perpendicular to the in-plane tip magnetization direction, respectively \cite{palotas18stt1}. Thus, the maximal STT (red) regions correspond to the out-of-plane spins outside of the skyrmionic objects and inside the skyrmionic cores as well as to the in-plane spins in the $\pm y$ direction perpendicular to $x$, and the minimal STT (blue) regions correspond to the maxima and minima of the charge current SP-STM maps (compare the top row of Fig.~\ref{Fig5} with the second row of Fig.~\ref{Fig2}). The topological properties of the skyrmionic spin textures are encoded in the number of minimal STT regions ($n_{\mathrm{STT_{min}}}$) obtained with any in-plane magnetized tip, i.e., $|Q|=n_{\mathrm{STT_{min}}}/2$, except for $Q=0$, for the same reason as for the charge current \cite{palotas17prb}.

The calculated 2D maps of the STT vectors and their components in Fig.~\ref{Fig5} show a large variety depending on the spin moment they are acting on ($T$ or $S$), the topology of the underlying skyrmionic spin structure, and the tunneling direction ($T\rightarrow S$ or $S\rightarrow T$). The change of the tip magnetization direction from out-of-plane (STT in Fig.~\ref{Fig4}) to in-plane reduces the overall symmetry of the coupled surface-tip system and affects the electron tunneling process, which is reflected by the lack of any obvious correspondence between the observed topological properties of the STT vector maps in Fig.~\ref{Fig5} in comparison with those of the skyrmionic textures in Fig.~\ref{Fig1}. For example, even though the $\mathbf{T}^{T\parallel}(V)$ maps in Fig.~\ref{Fig5} exhibit out-of-plane ($-z$) components outside the skyrmionic objects, the $\mathbf{T}^{T\parallel}$ vectors are restricted to be in the $yz$-plane at a tip magnetization direction of $x$. This clearly results in a lost connection between the topologies of the $\mathbf{T}^{T\parallel}$ maps and of the underlying spin structures.

\begin{figure*}[t]
\includegraphics[width=1.0\textwidth,angle=0]{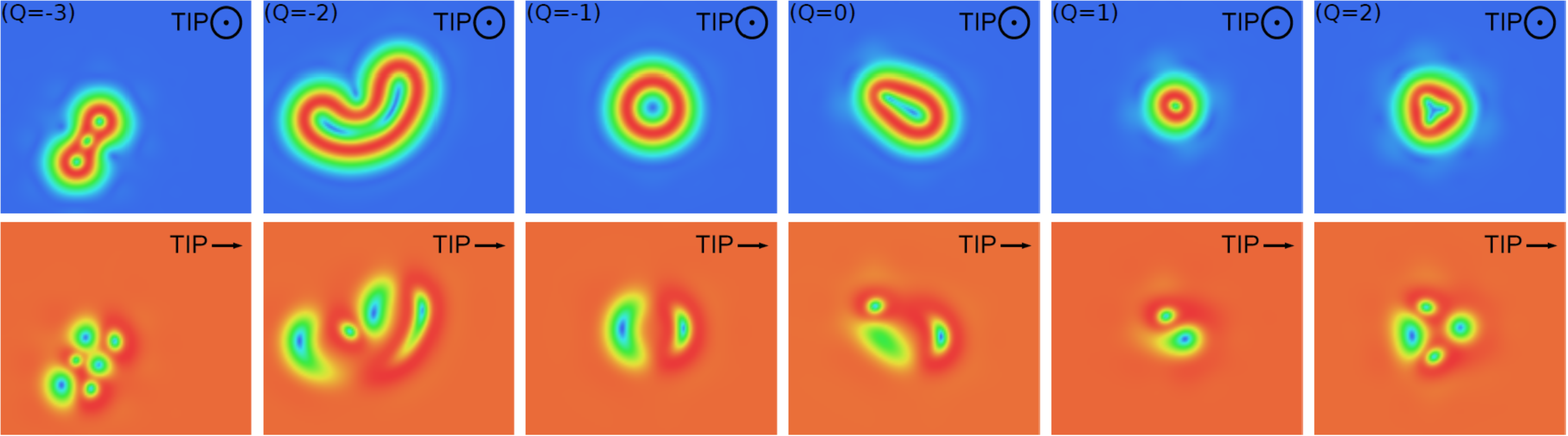}
\caption{\label{Fig6} STT efficiency ($\eta$) maps  based on Eq.~(\ref{Eq_STT_eff}) 6 \AA\;above the magnetic textures shown in Fig.~\ref{Fig1} at $|V|=1.5$ meV using an out-of-plane (first row) and an in-plane (second row) magnetized tip (magnetization directions are explicitly indicated). The color scales correspond to tip$+z$ (first row): red maximum at 23.2 meV/$\mu$A, blue minimum at 0 meV/$\mu$A; and tip$+x$ (second row): red maximum at 24.9 meV/$\mu$A, blue minimum at 1.6 meV/$\mu$A.}
\end{figure*}

The tunneling STT efficiency ($\eta$) is defined as the exerted absolute torque per current acting on the surface spins \cite{palotas16prb}, see Eq.~(\ref{Eq_STT_eff}). Next, we analyze the effect of the lateral position of the SP-STM tip on this quantity.

Figure \ref{Fig6} shows 2D maps of such calculated STT efficiencies for the topologically distinct skyrmions in Fig.~\ref{Fig1} at a constant-height condition, employing an out-of-plane and an in-plane magnetized scanning tip. For the out-of-plane ($+z$) magnetized tip (first row of Fig.~\ref{Fig6}) we obtain maximal $\eta$ values of 23.2 meV/$\mu$A ($\sim$0.9 $h/e$) and similar $\eta$ maps as for the STT magnitudes in the top row of Fig.~\ref{Fig4}. After a close inspection we find that the (red) regions of maximal $\eta$ are slightly shifted toward the core of the skyrmionic objects in all cases. The reason is the decreasing current values when moving from the rim toward the core of the skyrmionic structures at positive effective spin polarization ($P_SP_T>0$), see the top row of Fig.~\ref{Fig2}. This behavior would change to the opposite at negative effective spin polarization values ($P_SP_T<0$) \cite{palotas18stt1}, or at opposite tip polarity ($-z$). Therefore, independently of the exact skyrmionic topologies, the following lateral positions of the SP-STM tip are identified for achieving maximal STT efficiency when using out-of-plane magnetized tips: (i) the tip moved slightly \emph{toward} the core of the skyrmion from above the in-plane spins at $P_SP_T>0$ with $+z$-polarized tip or at $P_SP_T<0$ with $-z$-polarized tip, and (ii) the tip moved slightly \emph{outward} from above the in-plane spins at $P_SP_T<0$ with $+z$-polarized tip or at $P_SP_T>0$ with $-z$-polarized tip.

The maximal $\eta$ values are further increased to 24.9 meV/$\mu$A ($\sim$0.97 $h/e$) when considering an in-plane magnetized SP-STM tip. The obtained 2D maps of $\eta$ (second row of Fig.~\ref{Fig6}), again, resemble those of the STT magnitudes in Fig.~\ref{Fig5}. However, we observe a considerable asymmetry between minimum regions of $\eta$ (some blue-green regions are increased in size and the others are decreased in size with a red belt appearing around, exactly $|Q|$ numbers each, for $Q\ne0$) compared to those of the STT magnitudes. This is caused by the asymmetry of the charge current in these regions. We recall that the STT minima are obtained where the charge current has maxima or minima \cite{palotas18stt1}. For positive effective spin polarization ($P_SP_T>0$) the current is maximal (minimal) at tip positions above in-plane spins of the skyrmionic textures, which are parallel (antiparallel) magnetized with respect to the in-plane tip magnetization, see the second row of Fig.~\ref{Fig2} in comparison to Fig.~\ref{Fig1}. This behavior would change to the opposite at negative effective spin polarization values ($P_SP_T<0$) \cite{palotas18stt1}. Therefore, independently of the exact skyrmionic topologies, the following lateral positions of the SP-STM tip are identified for achieving maximal STT efficiency when using in-plane magnetized tips: the tip moved slightly around the in-plane spins, which are (i) \emph{antiparallel} to the tip magnetization at $P_SP_T>0$, and (ii) \emph{parallel} to the tip magnetization at $P_SP_T<0$.

To understand these results even better, a simple formula for the STT efficiency is proposed. Following Ref.~\onlinecite{palotas18stt1}, $\eta$ can be approximated by the dominating contribution from the surface atom $A$ closest to the tip apex atom, and from Eq.~(\ref{Eq_transport}) one obtains:
\begin{equation}
\eta_A(P_S,P_T,\phi_A)=\frac{|\mathbf{T}^{AS}|}{I_A}=\frac{h}{e}\frac{|P_{T}\sin\phi_A|\sqrt{1+P_{S}^2}}{1+P_{S}P_{T}\cos\phi_A}.
\label{Eq_STT_eff_A}
\end{equation}
This approximated STT efficiency has a maximal value of\\
$\eta_A^{\mathrm{max}}=(h/e)|P_{T}|\sqrt{[1+P_{S}^2]/[1-(P_{S}P_{T})^2]}$ at $\phi_A^{\mathrm{max}}=\arccos(-P_{S}P_{T})$. With the applied spin polarization parameters ($P_S=-0.5$ and $P_T=-0.8$) the above expression shows the following $\phi_A$-dependent function: $\eta_A(\phi_A)=(h/e)(0.8\sqrt{1.25}|\sin\phi_A|)/(1+0.4\cos\phi_A)$ that reaches a maximal value of 0.976 $h/e$ at $\phi_A^{\mathrm{max}}=113.58^\circ$. Thus, this simple $\eta_A$ expression can explain our numerically simulated maximal value of $\eta$ (0.97 $h/e$ with in-plane magnetized tips). The determined $\phi_A^{\mathrm{max}}$ angles at this maximum illustrate the extent of the necessary lateral tip movement with respect to the above indicated areas of spins for out-of-plane and in-plane magnetized SP-STM tips, in order to maximize the STT efficiency.

Finally, we note that the STT efficiency values are expected to decrease compared to the above reported values when realistic electron densities of states and orbital-dependent electron tunneling are accounted for \cite{palotas16prb}. Moreover, both the ferromagnetic core and the domain wall rim regions of the skyrmionic textures are expected to qualitatively exhibit the same electron charge and spin transport characteristics as identified above, independently of the size of the magnetic objects \cite{palotas18stt2}.

\section{Summary and conclusions}
\label{sec_conc}

Employing a combined electron charge and vector spin transport theory within spin-polarized scanning tunneling microscopy (SP-STM), the high-resolution tunneling electron spin transport properties of a set of topologically distinct magnetic skyrmionic textures were investigated on a surface of a 2D hexagonal lattice. The studied six prototypical (metastable) skyrmionic real-space spin structures possess various topological charges: $Q=-3,-2,-1,0,1,2$. We reported important insights into their spin transport properties and their topological relation to the spin textures by providing 2D maps of longitudinal spin current (LSC) and spin transfer torque (STT) magnitudes and vector quantities in high spatial resolution obtained by differently magnetized scanning tips. Using an out-of-plane magnetized tip it was found that the maps of the LSC vectors acting on the spins of the skyrmions and all STT vector components (out-of-plane, in-plane, and total STT) exhibit the same topology in the sense of the \emph{vorticity} as the real-space spin textures. In contrast, we found that an in-plane magnetized tip generally does not result in spin transport vector maps that are topologically equivalent to the underlying spin structure, except for the LSC vectors acting on the spins of the skyrmionic textures if $|P_T|<|P_S|$. For this relation of spin polarizations a topological equivalence between the LSC vectors acting on the spins of the skyrmions and the real-space spin textures in the sense of the \emph{topological charge} was identified independently of the magnetic orientation of the SP-STM tip. The magnitudes of the spin transport vector quantities exhibit close relations to charge current SP-STM images irrespectively of the skyrmionic topologies.

Moreover, we found that the STT efficiency acting on the spins of the skyrmions, $|\mathbf{T}^{S}|/I$, can reach large values up to $\sim$25 meV/$\mu$A ($\sim$0.97 $h/e$), and it considerably varies between large and small values depending on the lateral position of the SP-STM tip above the topological spin textures. We introduced a simple expression, Eq.~(\ref{Eq_STT_eff_A}), to explain the variation of the STT efficiency. Based on these results, and depending on the magnetic orientation of the tip and on the sign of the effective spin polarization of the magnetic tunnel junction, we identified lateral tip positions above the rim of the magnetic objects, where maximal STT efficiency can be achieved.

In this work we demonstrated the calculation of low-energy tunneling electron spin transport quantities in high spatial resolution above static topologically distinct skyrmionic spin structures. This will be extremely useful in the future in combination with atomistic spin dynamics methods in a dynamic setup involving higher energy tunneling electrons above fluctuating topological spin states, where the local STT vectors due to the presence of an SP-STM tip can be calculated following our model, and thermal effects can concomitantly be included. Furthermore, spin dynamics simulations are expected to reveal the relationship between the regions of maximal STT efficiency identified in our present work, and the lateral tip positions where the dynamical process of the switching can really be optimized. Such a combination would result in detailed microscopic insights into the creation, movement, and annihilation of topologically distinct surface magnetic skyrmionic textures by the magnetic STM tip, complementing existing minimum-energy-path-based methods.

\section{Acknowledgments}

Financial supports of the National Research, Development, and Innovation Office of Hungary (NKFIH) under Projects No.\ K115575, No.\ PD120917, No.\ FK124100, and No.\ K131938, of the BME Nanotechnology and Materials Science TKP2020 IE grant of NKFIH Hungary (BME IE-NAT TKP2020), of the SASPRO Fellowship of the Slovak Academy of Sciences (project No.\ 1239/02/01), and of the Alexander von Humboldt Foundation are gratefully acknowledged.

\end{document}